\documentclass[final,5p,times,twocolumn]{elsarticle}
\usepackage{graphicx}
\usepackage{color}
\begin{document}
\begin{frontmatter}
\title{Current in nanojunctions : Effects of reservoir coupling}
\author{Hari Kumar Yadalam}
\author{Upendra Harbola}%
\address{Department of Inorganic and Physical Chemistry, Indian Institute of Science, Bangalore 560012, India.}
\begin{abstract}
We study the effect of system reservoir coupling on currents 
flowing through quantum  junctions. 
We consider two simple double-quantum dot configurations coupled to two 
external fermionic reservoirs 
and study the net current flowing between 
the two reservoirs. The net current is partitioned into currents carried by the 
eigenstates of the system and by the coherences 
between the eigenstates induced due to coupling with the reservoirs. We find that 
current carried by populations is always positive whereas current carried by 
coherences are negative for large couplings. This results in a non-monotonic dependence of the 
net current on the coupling strength. We find that in certain cases, the net current can 
vanish at large couplings due to cancellation
between currents carried by the eigenstates and by the coherences.  
These results provide new insights 
into the non-trivial role of  system-reservoir couplings on electron transport 
through quantum dot junctions.
In the presence of weak 
coulomb interactions, net current as a function of system reservoir coupling strength shows similar 
trends as for the non-interacting case.
\end{abstract}
\end{frontmatter}
 
\section{Introduction}
\label{introduction}

\noindent
Transport properties of  quantum  junctions have been studied for 
over two decades motivated not only by their technological relevance but also 
the opportunities they provide to explore fundamental physics. For example 
quantum dot  junctions provide a good platform for verification of 
fundamental concepts, like fluctuation theorems \cite{Esposito2009,Utsumi2010} 
. 
There have also been a lot of technologically 
relevant proposals of diodes \cite{Aviram1974}, transistors 
\cite{Cardamone2006}, heat engines 
\cite{Scully2003,Goswami2013}, which can be 
realized using quantum junctions made of single molecules or quantum dots. 
Quantum dot junctions can also serve as promising 
candidates for realizing quantum computers \cite{Loss1998}. 

Current flowing through quantum dot 
junctions \cite{Van2002} and molecular junctions \cite{Tao2006,Xiang2016,Baldea2016,Cuevas2010,Cuniberti2006}
have been measured experimentally and studied  using various theoretical formulations like quantum master equations (QME) 
\cite{Harbola2006}, scattering matrix (SM) \cite{Datta1997}, and non equilibrium 
Green's function (NEGF) method \cite{Haug2008}. QME and SM approaches are 
valid within a certain parameter regime, but NEGF method is exact and  
can be applied in all regimes, although analytically tractable results can be 
obtained only for non-interacting systems.

Although a good amount of theoretical work  on quantum conduction exists in the 
literature \cite{Datta1997},  
however the role of system-reservoir coupling  
has not been explored much, except for few works. For example in experiments 
performed with carbon nanotube junctions reported in Ref.\cite{Chiu2007}, the 
importance of non-point like contact of reservoir system coupling was observed. 
The effect of finite contact length was studied in Ref. \cite{Nemec2008} 
using tight binding models, and it was demonstrated that the transmission can be 
enhanced at lower system reservoir coupling 
strengths by increasing the contact length.
Further, in Ref.\cite{Shahbazyan1994}, the effect of reservoir induced coupling between 
quantum dots on the current was studied. It is important to note here that, the system-reservoir coupling strength 
can be tuned using external gate potentials in quantum dot junctions \cite{Van2002} and 
can be tuned in molecular junctions \cite{Su2016} by tuning the density of states of metal near fermi-energy \cite{Seminario2001,Beebe2002,Adak2015},
by tuning orbital overlaps of metal and molecule \cite{Ko2009} or by  chemical gating \cite{Danilov2008}.

To gain more understanding on the role of system reservoir coupling 
strength, we ask the question, "How does the current vary as system-reservoir 
coupling is changed?". To answer this question, 
we note that, in a simple scattering picture, the system-reservoir coupling 
offers (contact) resistance to the 
tunneling electrons. Within the quantum master equation 
formulation (Lindblad quantum master equation), the current increases  
monotonically as the coupling is increased.
However, this does not present the complete picture and it is not at all obvious 
what happens as 
one goes beyond the regime of QME or simple scattering picture.  
In Ref.\cite{Gurvitz1996}, scattering formalism under weak reservoir 
coupling was used to study the effect of reservoir induced coherences on the 
net current through a coupled double-quantum dot model. 

In this work we explore the effect of strong system-reservoir couplings on the 
net current 
flowing through quantum junctions using NEGF formulation. 
The advantages of the NEGF formulation over the other formulations discussed
above are two folds. First,  in most practical cases, it provides an exact method to 
compute the current in molecular junctions. Secondly, this is a standard well established 
method to include effects arising due to many-body interactions, as we shall discuss in the later
part of this paper.

In the following, we find that the net current 
is not always an increasing function of the coupling strength. In fact, 
surprisingly, 
we find that for certain cases the net current may diminish at large coupling 
strengths. 
As we discuss below, this surprising behavior is a consequence of the quantum 
interference
between the eigenstates which carries a negative (against the applied bias) 
current that may
cancel the currents coming from the eigenstates. 
In the absence of the interferences, the net current always shows a monotonic
increase with the reservoir couplings.
The current  behavior for large couplings, of course,  depends on the quantum 
dot configuration and is not universal. 
For certain configurations, there is an optimal value of the coupling strength 
at which the current 
is maximal. A similar non-monotonic behavior of heat current in spin-boson
model\cite{Velizhanin2008,Agarwalla2017} 
and of energy flux through externally driven molecular junction\cite{White2013} 
has been observed. 

Recently the effect of quantum interference on current flowing through molecular junctions has been studied 
experimentally \cite{Aradhya2012,Guedon2012,Arroyo2013,Vazquez2012} and theoretically 
\cite{Lambert2015,Markussen2010,Solomon2008,Hansen2009,Reuter2014,Andrews2008}. 
Various device proposals making use of quantum interference effects have been made, see for example quantum transistor \cite{Cardamone2006}, 
thermoelectric engines \cite{Chen2017,Nozaki2014,Lambert2016}, molecular switch \cite{Baer2002}.
It was shown both experimentally and theoretically that vibrations suppress destructive interference effect leading to 
enhancement of current \cite{Hartle2011,Ballmann2012}.
In a recent work by Markussen and Thygesen \cite{Markussen2014},
the effect of temperature on the junction conductance has been studied using 
interacting quantum dot model. The interference effect was shown to lead to stronger temperature dependence of conductance. 
In the present work, however, we focus on interference 
effects in the strong molecule-reservoir coupling regime where 
such effects are significant and play a crucial role, as discussed below. 
We find that the temperature (broadening of fermi functions) only makes small quantitative changes, the qualitative
behavior of the junction conductance remains the same.

To explore this current behavior we 
consider two simple (non-interacting) models both consist of two quantum dots coupled to two 
fermionic reservoirs  but differ in their configurations. This is discussed in 
the next section.

\section{Model Hamiltonian and current calculation}
\begin{figure}[h]
\centering
\includegraphics[width=7.8cm,height=2.4cm]{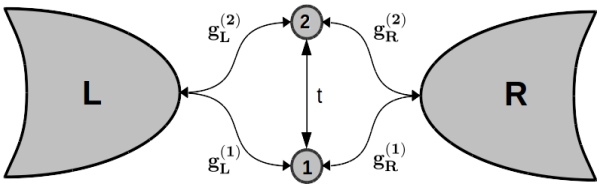}
\caption{Schematic of the model system considered. It consists
of two localized sites coupled to two fermionic reservoirs,
left ($L$) and right ($R$). $g_\alpha^{(x)}$
 is the strength of the coupling between
the $\alpha^{th}$ reservoir and the $x^{th}$ site and $t$ is the inter-site
coupling strength.}
\label{fig-1}
\end{figure}

 We start by considering a simple model shown in Fig. \ref{fig-1}. 
It consists of two quantum dots each having a single electron orbital coupled to 
each other and also coupled to two 
fermionic reservoirs.

The Hamiltonian describing this model is given as, 
\begin{eqnarray}
\label{eq-1}
 \hat{H}&=&\sum_{i,j=1}^2 H_{0_{ij}}^{} c_i^{\dag} 
c_j^{}+\mathop{\sum_k}_{\alpha=L,R}\epsilon_{\alpha,k}^{} d_{\alpha k}^\dag 
d_{\alpha k}^{}\nonumber\\
         &+&\sum_k\left[g_L^{(1)} d_{Lk}^\dag c_1^{}+g_R^{(1)} d_{Rk}^\dag 
c_1^{}+g_L^{(2)} d_{Lk}^\dag c_2^{}\right.\nonumber\\
&+& \left. g_R^{(2)} d_{Rk}^\dag c_2^{}+h.c.\right]
\end{eqnarray}
where
\begin{eqnarray}
\label{eq-2}
 H_{0}= \left(
\begin{array}{cc}
          \epsilon_1 & -t\\
          -t & \epsilon_2\\
\end{array}
\right)
\end{eqnarray}
is the single particle Hamiltonian for the isolated molecule. 
Here $c_{i}$ ($c_{i}^{\dag}$) are the fermionic annihilation (creation) 
operators 
for destroying (creating) electron at site '$i$' and similarly $d_{\alpha k}$ 
($d_{\alpha k}^{\dag}$) are operators for 
destroying (creating) electron in state labeled by 'k' in the '$\alpha$' 
reservoir
($\alpha=L/R$). First two terms in the Hamiltonian represent isolated system and 
reservoir Hamiltonians, and the third term represents hybridization 
between system and reservoirs with $g_\alpha^{(1)}$ and $g_\alpha^{(2)}$ 
representing 
coupling of the $\alpha$th reservoir with dot $(1)$ and dot $(2)$, respectively.
We have also assumed wide-band approximation (system-reservoir coupling is 
independent of 'k').

The net current $I_L$ flowing into the left reservoir 
is given by the rate of change of charge on the left reservoir, i.e.,
$I_L(t)=\frac{d}{dt}\langle-e\sum_k d_{Lk}^\dag d_{Lk}\rangle$. The net current 
can be expressed in terms of system greater and lesser Green's functions 
\cite{Meir1992,Haug2008}  $G^{>/<}$ as ,
\begin{eqnarray}
\label{eq-8}
I_L=\frac{e}{h}\int_{-\infty}^{+\infty} d\omega Tr\big[\Sigma^{<}_{L}(\omega)
G^{>}(\omega)-G^{<}(\omega)\Sigma^{>}_{L}(\omega)\big], 
\end{eqnarray}
where $\Sigma^{>/<}_{L}(\omega)$ and $G^{>/<}(\omega)$ are 
(energy domain) Fourier transformed greater and lesser projections 
of contour-ordered self-energy due to left reservoir and the system Greens' 
functions \cite{app}. A similar expression, obtained by replacing 
$L\Leftrightarrow R$ in Eq. (\ref{eq-8}), holds true 
for the right current, $I_R$. At steady state the left and the right 
currents must be the same in magnitude, $|I_R|=|I_L|$, which is referred to
as the left-right symmetry at steady state.

The Green's functions obtained by solving equation of motion\cite{app} in energy domain can be used in Eq. (\ref{eq-8}) to get 
expression for the net current, $I_{L}$ \cite{Haug2008,Yadalam2016}.

For simplicity we consider two cases : serially coupled dot system (obtained in 
the limit $g_L^{(2)}=g_R^{(1)}=0$) and side coupled dot system 
(obtained in the limit $g_L^{(2)}=g_R^{(2)}=0$). We further assume that site 
energies are same as the Fermi energies of the two reservoirs (set to zero).

The net current for serially coupled dot system is obtained as (we use units 
such that $e=1$, and $h=1$)
\begin{eqnarray}
 I_L&=&\int_{-\infty}^{+\infty}d\omega\big[\frac{\Gamma^2 
t^2}{(\omega^2-t^2-(\frac{\Gamma}{2})^2)^2+\omega^2 
\Gamma^2}\big][f_L(\omega)-f_R(\omega)].\nonumber\\
\end{eqnarray}
For the side coupled dot system, the net current is
\begin{eqnarray}
  I_L&=&\int_{-\infty}^{+\infty}d\omega\big[\frac{\Gamma^2 
\omega^2}{(\omega^2-t^2)^2+\omega^2 
\Gamma^2}\big][f_L(\omega)-f_R(\omega)].\nonumber\\
\end{eqnarray}
Here $f_{\alpha}(\omega)=\frac{1}{e^{\beta_{\alpha}(\omega-\mu_{\alpha})}+1}$ is 
the Fermi function of two reservoirs 
($\alpha=1,2$), $\beta_{\alpha}$ and $\mu_{\alpha}$ are, respectively, the 
inverse temperature and 
the chemical potential of $\alpha^{th}$ reservoir. Here we assumed that all the 
relevant non-zero couplings to the reservoirs 
are identical,  $g_\alpha^{(x)}=g$. The coupling strength $\Gamma=2\pi \rho 
|g|^2$, where $\rho$ is the density of states of the reservoirs assumed to be 
energy independent and identical for both the reservoirs.

\begin{figure}[!htbp]
\centering
\includegraphics[width=7.2cm,height=5.0cm]{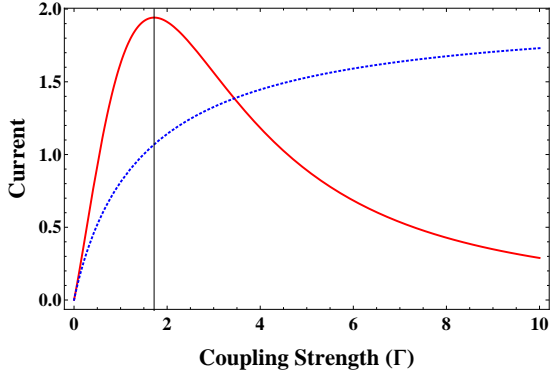}
\caption{ (Color online) Net current as a function of $\Gamma$ for 
serially coupled double quantum dot system (red-continuous) and side coupled 
double quantum dot system(blue-dotted) systems. 
Here $\beta_L^{-1}=\beta_R^{-1}=0$, $\mu_L=\frac{V}{2}$ and 
$\mu_R=-\frac{V}{2}$ with $V=2$.  All energy scales are in units of $t$ and 
current is in units of $et/h$ with $t=1$.}
\label{fig-2}
\end{figure}

Currents in Eq.(7) and Eq.(8) are plotted against $\Gamma$ in Fig. (\ref{fig-2}) 
for $t=1$, $\mu_L=1$, $\mu_R=-1$ and $\beta_L^{-1}=\beta_R^{-1}=0$.  
Throughout this work, all the energy scales and currents 
are expressed in units of $t$ and $\frac{e}{h}t$, respectively.
It is clear that the net current is not always an increasing function of $\Gamma$. 
For side coupled dot system, net current is an increasing function of $\Gamma$  
and 
saturates asymptotically to a constant value for large $\Gamma$, which for zero 
temperature is simply proportional to the difference in chemical potentials of 
the two reservoirs. However, for serial coupled dot system, the net current 
shows a non-monotonic behavior and settles to zero for large $\Gamma$. The 
latter case is very counter intuitive and, in order to understand these
two completely different current behaviors, below we analyze currents in the 
eigenbasis of the system.
\section{Partitioning the current}
\label{sec-3} 
We define a unitary transformation matrix, $\mathcal{U}$, which diagonalizes the 
system Hamiltonian $H_{0}^{}$, i.e., 
$\mathcal{U}=\frac{1}{\sqrt{2}}\left(\begin{array}{cc} 1&1\\1&-1 \end{array}\right)$. We next 
transform lesser and greater Green's functions $G^{</>}(\omega)$ and lesser and 
greater left-reservoir self-energies $\Sigma_{L}^{</>}(\omega)$ 
into the eigenbasis using, $A\rightarrow\bar{A}=\mathcal{U}^{\dag}A\mathcal{U}$, 
where 
$A$ is any matrix defined in the local basis.
Thus transforming Eq. (\ref{eq-8}) to eigenbasis, the net current can be 
partitioned into currents carried by the population in the bonding state 
($I_{b}$),  
population in the anti-bonding state ($I_{a}$), and the current carried by 
coherences between these two states ($I_{c}$). 
The expressions for individual contributions are given as 
\begin{eqnarray}
\label{Ib}
I_{b}&=&\int_{-\infty}^{+\infty}d\omega\big[{\Sigma^{<}_{L}}_{bb}
(\omega)G^{>}_{bb}(\omega)-G^{<}_{bb}(\omega){\Sigma^{>}_{L}}_{bb}(\omega)\big],
\end{eqnarray}
\begin{eqnarray}
\label{Ia}
I_{a}&=&\int_{-\infty}^{+\infty}d\omega\big[{\Sigma^{<}_{L}}_{aa}
(\omega)G^{>}_{aa}(\omega)-G^{<}_{aa}(\omega){\Sigma^{>}_{L}}_{aa}(\omega)\big]
\end{eqnarray}
and
\begin{eqnarray}
\label{Ic}
I_{c}&=&\int_{-\infty}^{+\infty}d\omega\big[{\Sigma^{<}_{L}}_{ba}
(\omega)G^{>}_{ab}(\omega)-G^{<}_{ab}(\omega){\Sigma^{>}_{L}}_{ba}
(\omega)\nonumber\\
&&+{\Sigma^{<}_{L}}_{ab}(\omega)G^{>}_{ba}(\omega)-G^{<}_{ba}(\omega){\Sigma^{>}
_{L}}_{ab}(\omega)\big].
\end{eqnarray}
Partitioning of the net current in Eqs. (\ref{Ib})- (\ref{Ic}) is based on
the fact that the Greens' functions $G^<_{xx}$ and $G^>_{xx}$, where $x=a,b$, correspond 
to the population of state $x$, while $G^<_{xy}$ and $G^>_{xy}$ give coherences between
the states $x$ and $y$. Similarly $I_R$ can also be partitioned
in terms of currents carried by the populations and the coherences. It is straightforward to
show that these currents are individually conserved, i.e, left-right symmetry holds 
for each current.

We next specialize to two simple models introduced in the previous section to 
gain a better insight into the role of 
system reservoir coupling strength on the current.

\begin{figure*}[tbh]
\centering
\begin{tabular}{ccc}
\includegraphics[width=7.2cm,height=5.0cm]{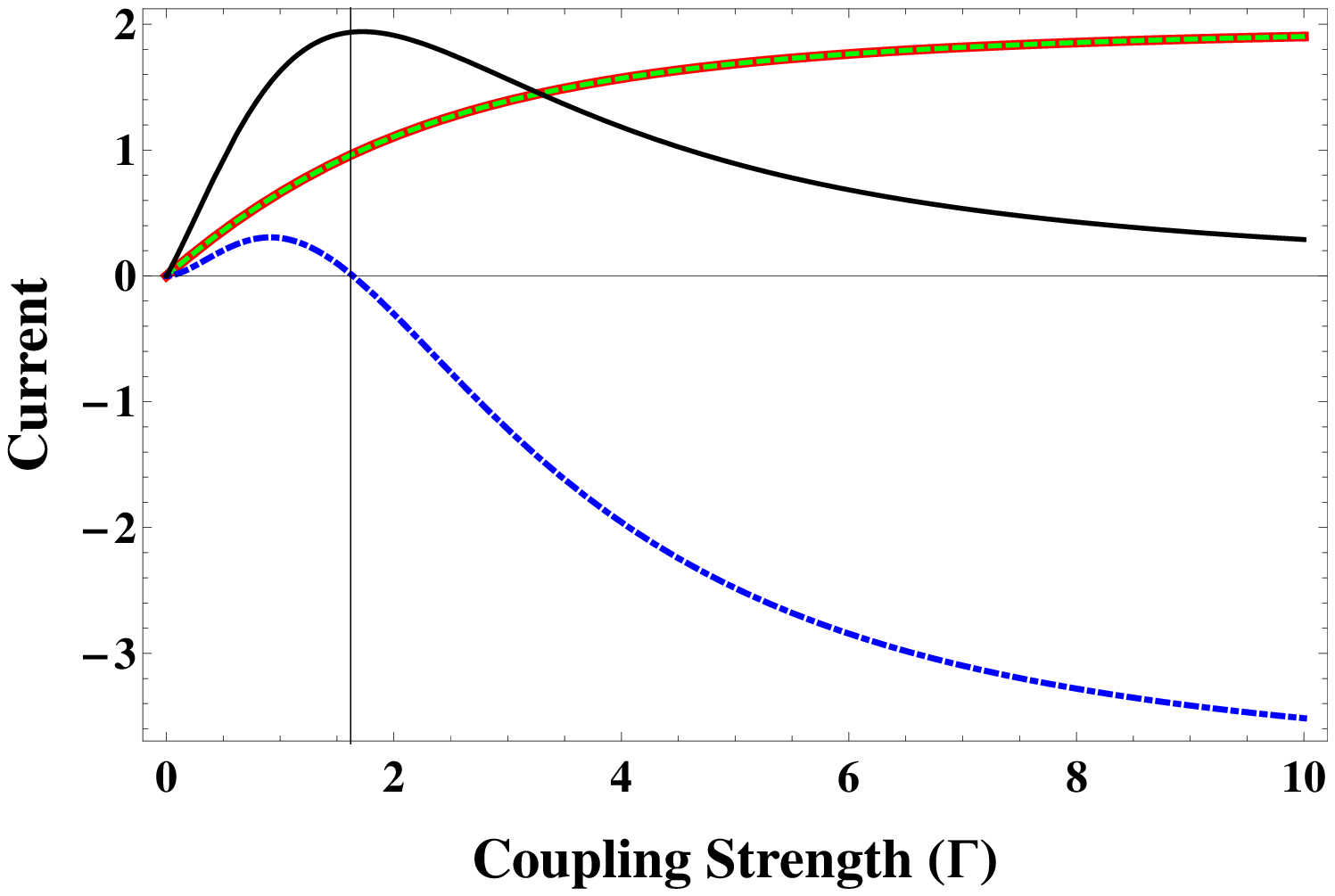} &
\includegraphics[width=7.2cm,height=5.0cm]{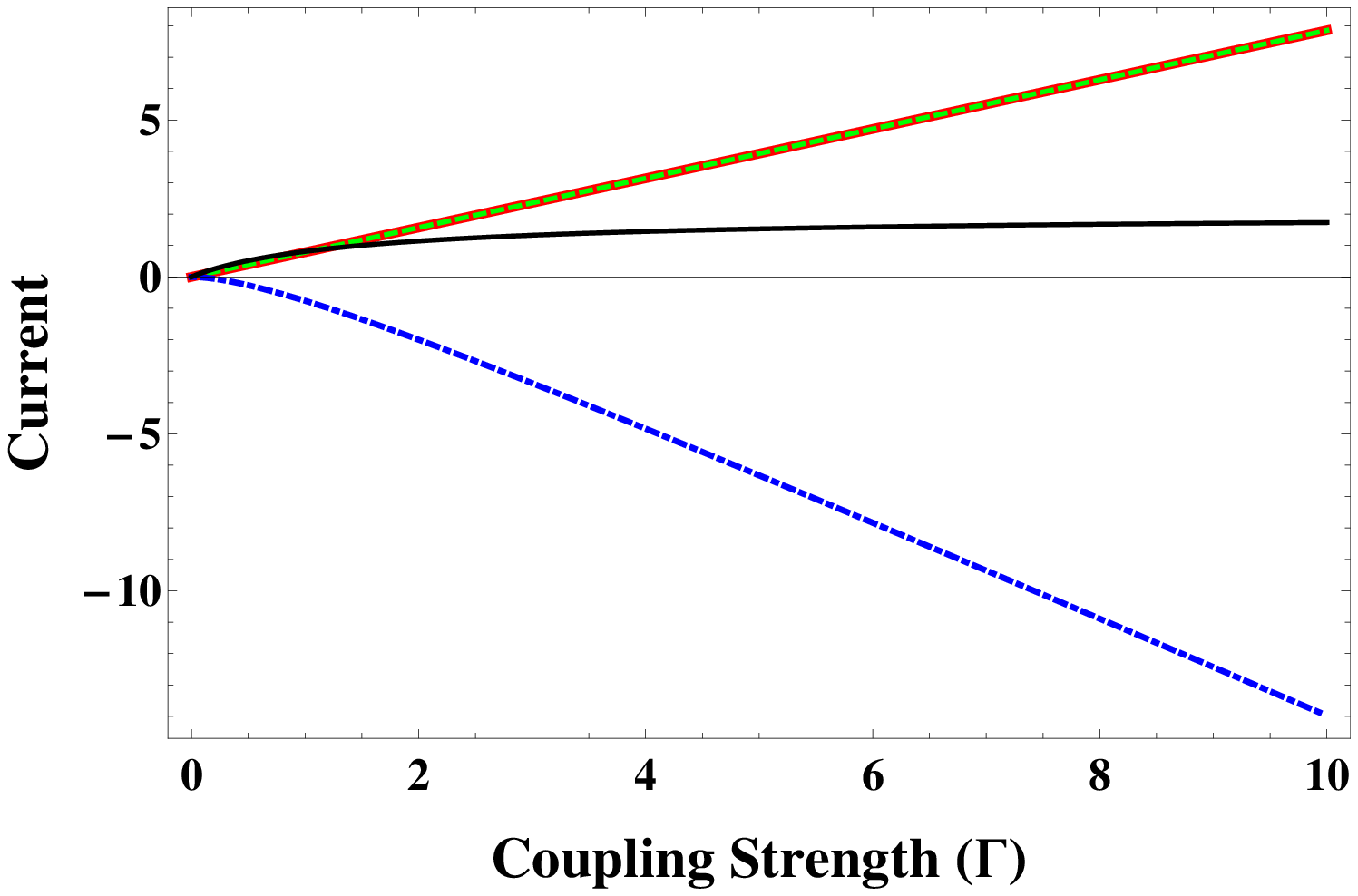} 
\end{tabular}
\caption{ (Color online) Currents carried by population in bonding (red-thick) state, population in 
anti-bonding (green-dashed) state and coherences between the bonding and anti-bonding states (blue-dot-dashed) 
along with the net current (black-thin) 
 for serially coupled (left) and side coupled (right) double quantum dot system as a function of $\Gamma$ with 
 all parameters being same as in Fig. (\ref{fig-2}).}
\label{fig-3}
\end{figure*}

\textbf{Serially coupled system :}
For the serially coupled double quantum dot system, explicit expressions for 
$I_{b}$, $I_{a}$ and $I_{c}$ are
\begin{eqnarray}
\label{Ib-1}
I_{b}&=&\int_{-\infty}^{+\infty}d\omega\big[\frac{(\frac{\Gamma}{2})^2}{
(\omega+t)^2+(\frac{\Gamma}{2})^2}\big][f_L(\omega)-f_R(\omega)],\nonumber\\
\end{eqnarray}
\begin{eqnarray}
\label{Ia-1}
I_{a}&=&\int_{-\infty}^{+\infty}d\omega\big[\frac{(\frac{\Gamma}{2})^2}{
(\omega-t)^2+(\frac{\Gamma}{2})^2}\big][f_L(\omega)-f_R(\omega)]\nonumber\\
\end{eqnarray}
and
\begin{eqnarray}
\label{Ic-1}
I_{c}&=&\int_{-\infty}^{+\infty}d\omega\big[\frac{-2  (\frac{\Gamma}{2})^2 
\left(\omega^2 - t^2 + 
(\frac{\Gamma}{2})^2\right)}{(\omega^2-t^2-(\frac{\Gamma}{2})^2)^2+\omega^2 
\Gamma^2}\big][f_L(\omega)-f_R(\omega)].\nonumber\\
\end{eqnarray}
The above integrals can be easily performed for zero temperature case 
($\beta_L^{-1}=\beta_R^{-1}=0$) to get 
\begin{eqnarray}
\label{Ib0-1}
I_{b}&=&\frac{\Gamma}{2}
\left[\tan^{-1}(\frac{V+2t}{\Gamma})+\tan^{-1}(\frac{V-2t}{\Gamma})\right]
\end{eqnarray}
and $I_a=I_b$ at zero temperature for the symmetrically biased system 
($\mu_L=\frac{V}{2}$ and $\mu_R=-\frac{V}{2}$). The coherent contribution is 
obtained as
\begin{eqnarray}
\label{Ic0-1}
I_{c}&=&\left(\frac{\Gamma^2}{4t^2+\Gamma^2}\right)\Bigg\{ 
t\log\left(\frac{(V+2t)^2+\Gamma^2}{(V-2t)^2+\Gamma^2}\right)\nonumber\\
&-&\Gamma\left[\tan^{-1}\left(\frac{V+2t}{\Gamma}\right)+\tan^{-1}\left(\frac{
V-2t}{\Gamma}\right)\right]\Bigg\}
\end{eqnarray}

The expressions for currents from the bonding and the anti-bonding 
orbitals, Eqs. (\ref{Ib-1}) and (\ref{Ia-1}), are identical to the one obtained for
a single resonant level with energies $-t$ and $t$, respectively \cite{Haug2008}. 
These contributions are always positive (throughout, we assume $\mu_L>\mu_R$ and the 
two reservoirs have the same temperature).   However, as noted from Eq. 
(\ref{Ic-1}) or
(\ref{Ic0-1}), the  coherent contribution can be positive or negative depending 
on the relative values of the coupling strengths $\Gamma$ and $t$. 
For large $\Gamma$, the logarithmic term vanishes and coherent contribution is 
always negative which can compete with the contributions from the
populations.  
Current contributions from eigenstate populations and coherences together with 
the net current are plotted as a function of $\Gamma$ in the left panel of Fig. (\ref{fig-3}). 
Here bonding and anti-bonding (population) contributions 
are equal due to the parameters chosen ($\epsilon_1=\epsilon_2=0$ and 
$\mu_R=-\mu_L$). These contributions increase with $\Gamma$ and saturate to a 
non zero constant 
value for large $\Gamma$, which corresponds to unit conductance $(e^2/h)$ per 
electron channel. The coherent contribution shows non monotonic trend, initially 
increases but finally settles down to a negative value which is equal to the sum 
of bonding and anti-bonding contributions for large $\Gamma$. This non-monotonic 
character in $I_c$  is seen only for bias values $V\leq 2t$. For large values of 
the bias $V>>2t$, the coherent contribution is always negative. Thus for 
intermediate bias values, it should be possible to maximize the net current by 
suitably choosing the coupling strength. For $V<<2t$, the coherent contribution 
vanishes if $\Gamma=2t$ and the net current is maximum.
For large $\Gamma$, the conductivity of the two population channels (bonding and 
anti-bonding states) is unity (in units of $\frac{e^2}{h}$) while that of the 
coherent channel approaches to $2$, although in the opposite direction to the 
applied bias. Thus for large $\Gamma$, both population channels and coherence 
channel conduct equal current  but in the opposite directions, which leads to  a 
vanishing net current for large $\Gamma$.

\textbf{Side coupled system :}
Next we consider the case when $g_L^{(2)}=g_R^{(2)}=0$.
In this case the explicit expressions for $I_{b}$, $I_{a}$ and $I_{c}$ are 
obtained as follows.
\begin{eqnarray}
I_{b}&=&\frac{1}{4}\int_{-\infty}^{+\infty}d\omega\big[\frac{\Gamma ^2 
(\omega-t)^2}{(\omega^2-t^2)^2+\omega^2\Gamma^2}\big][f_L(\omega )-f_R(\omega 
)],\nonumber\\
\end{eqnarray}
\begin{eqnarray}
I_{a}&=&\frac{1}{4}\int_{-\infty}^{+\infty}d\omega\big[\frac{\Gamma ^2 
(\omega+t)^2}{(\omega^2-t^2)^2+\omega^2\Gamma^2}\big][f_L(\omega )-f_R(\omega 
)]\nonumber\\
\end{eqnarray}
and
\begin{eqnarray}
I_{c}&=&\frac{1}{2}\int_{-\infty}^{+\infty}d\omega\big[\frac{\Gamma ^2 
(\omega^2-t^2)}{(\omega^2-t^2)^2+\omega^2\Gamma^2}\big][f_L(\omega )-f_R(\omega 
)].\nonumber\\
\end{eqnarray}
The analytic expressions for these currents for zero temperature case are 
given by,
\begin{eqnarray}
\label{Ib-sc}
I_{b}&=&\big(\frac{\Gamma}{2}\big)^2\Bigg\{\frac{(a_1-t)^2}{
(a_1-a_2)(a_1-a_3)(a_1-a_4)}\log(\frac{a_1-\frac{V}{2}}{a_1+\frac{V}{2}}
)\nonumber\\
&&+\frac{(a_2-t)^2}{(a_2-a_1)(a_2-a_3)(a_2-a_4)}\log(\frac{a_2-\frac{V}{2}}{
a_2+\frac{V}{2}})\nonumber\\
&&+\frac{(a_3-t)^2}{(a_3-a_1)(a_3-a_2)(a_3-a_4)}\log(\frac{a_3-\frac{V}{2}}{
a_3+\frac{V}{2}})\nonumber\\
&&+\frac{(a_4-t)^2}{(a_4-a_1)(a_4-a_2)(a_4-a_3)}\log(\frac{a_4-\frac{V}{2}}{
a_4+\frac{V}{2}})\Bigg\}.
\end{eqnarray}
$I_a$ is obtained by replacing $t$ with $-t$ in  Eq. (\ref{Ib-sc}), and
\begin{eqnarray}
I_{c}&=&\big(\frac{\Gamma}{2}\big)^2\Bigg\{\frac{1}{(a_1-a_2)}\Big[\log(\frac{
a_1-\frac{V}{2}}{a_1+\frac{V}{2}})-\log(\frac{a_2-\frac{V}{2}}{a_2+\frac{V}{2}}
)\Big]\nonumber\\
&&+\frac{1}{(a_3-a_4)}\Big[\log(\frac{a_3-\frac{V}{2}}{a_3+\frac{V}{2}}
)-\log(\frac{a_4-\frac{V}{2}}{a_4+\frac{V}{2}})\Big]\Bigg\}
\end{eqnarray}
where $a_1=-i\frac{\Gamma}{2}+\sqrt{t^2-(\frac{\Gamma}{2})^2}$, 
$a_2=-i\frac{\Gamma}{2}-\sqrt{t^2-(\frac{\Gamma}{2})^2}$, 
$a_3=i\frac{\Gamma}{2}+\sqrt{t^2-(\frac{\Gamma}{2})^2}$ and 
$a_4=i\frac{\Gamma}{2}-\sqrt{t^2-(\frac{\Gamma}{2})^2}$.

For $\Gamma\gg t$, the current contributions, $I_b$ and $I_c$, acquire the 
simple form,
\begin{eqnarray}
\label{Ib-n-Ic}
I_b&=& \frac{\Gamma}{2}\left[\mbox{tan}^{-1}\left(\frac{\Gamma 
V}{2t^2}\right)+\mbox{tan}^{-1}\left(\frac{V}{2\Gamma}\right)\right]\nonumber\\
I_c&=& \Gamma 
\left[\mbox{tan}^{-1}\left(\frac{V}{2\Gamma}\right)-\mbox{tan}^{-1}\left(\frac{
\Gamma V}{2t^2}\right)\right].
\end{eqnarray}
Unlike the serially coupled case, in this case both contributions, population as 
well as the coherences, grow linearly with $\Gamma$. However, their sum, the 
total current, saturates to the value $\Gamma{\mbox{tan}^{-1}(V/2\Gamma)}$. We 
again notice that contributions from the bonding and the anti-bonding states are 
always positive while the coherent 
contribution is always negative for large $\Gamma$. This is shown in the right panel of Fig. 
(\ref{fig-3}). The rate of increase of the currents though the eigenstates
is precisely half of the rate with which current increases (in the opposite 
direction) via the coherences. Thus the net rate is zero and the total current
saturates to a constant value.

In this section, we have derived some analytical results for simple 
noninteracting model systems to study the effects of system-reservoir
coupling on the net current. A natural question arises as to the validity of this
result in more realistic systems. To check this, in the following section, we introduce 
electron-electron interaction in the system. However it becomes difficult to obtain analytic expressions for currents, 
hence we present results based on numerical calculations.

\section{Effect of coulomb interaction}
\label{sec-3a} 
To explore effect of coulomb interaction on the trend observed above, we add the 
following interaction part to the system Hamiltonian,
\begin{eqnarray} 
H_{int}&=&\frac{1}{2}\sum_{i,j=1,2}V_{ij}^{}c_{i}^{\dag}
c_{i}^{}c_{j}^{\dag}c_{j}
\end{eqnarray}
where $V_{ij}^{}=U(1-\delta_{ij}^{})$. This
(Coulomb) interaction leads to an extra self-energy in the equation of motion of 
the Greens' function\cite{app}. 
We compute this self-energy within the  Hartree-Fock (HF) or mean-field approximation 
and $GW$ approximation\cite{Hedin1965,Harbola2006gw,Thygesen2008,Spataru2009,Stefanucci2013}. 
 It has been shown in Ref.\cite{Tandetzky2015} that within $GW$ 
approximation one can have multiple solutions.
However, in the weak interaction limit, $(U/\Gamma)\leq 1$, there is only one 
unique physical solution.

\begin{figure*}[tbh]
\centering
\begin{tabular}{ccc}
\includegraphics[width=7.2cm,height=5.0cm]{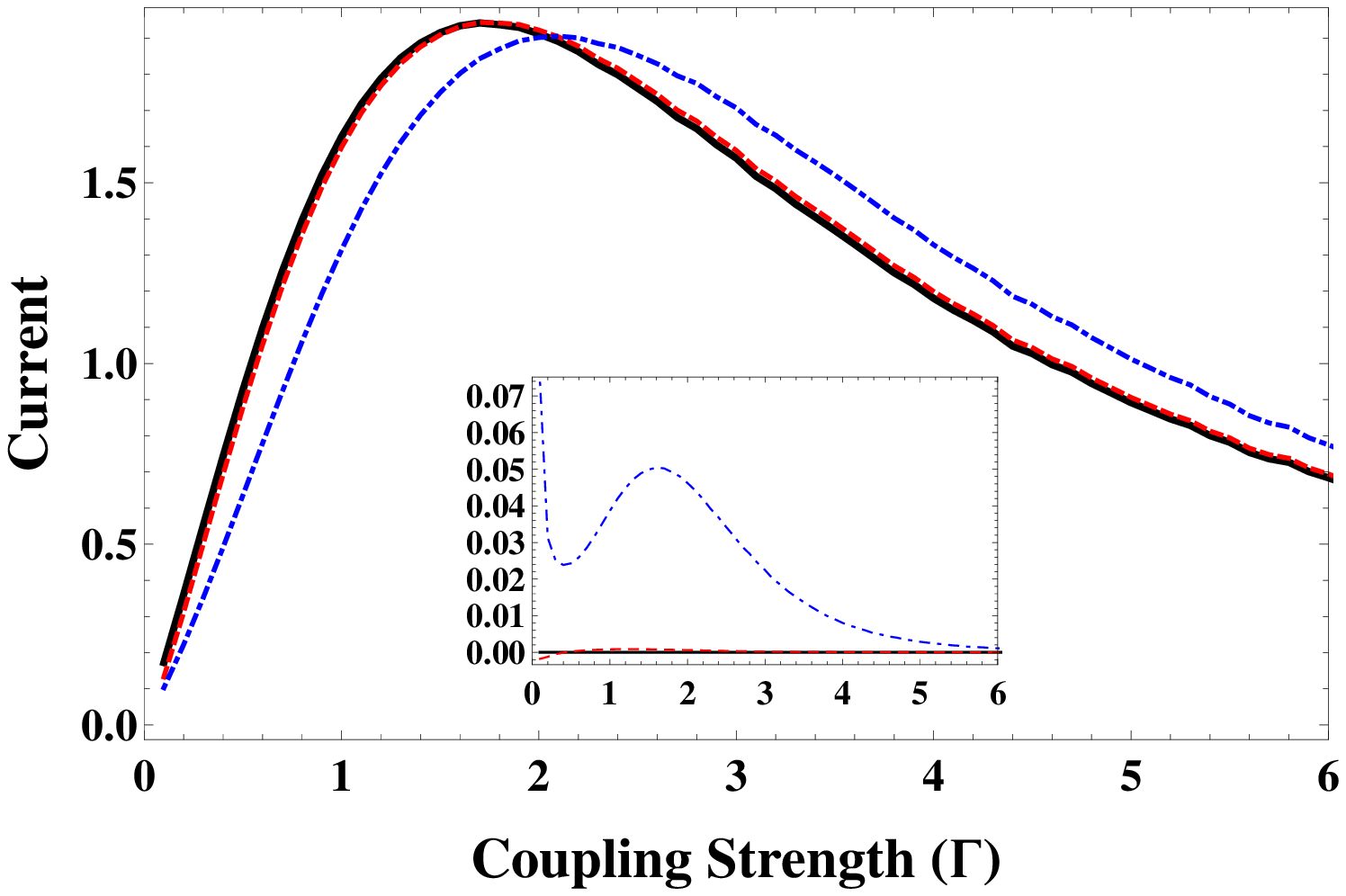} &
\includegraphics[width=7.2cm,height=5.0cm]{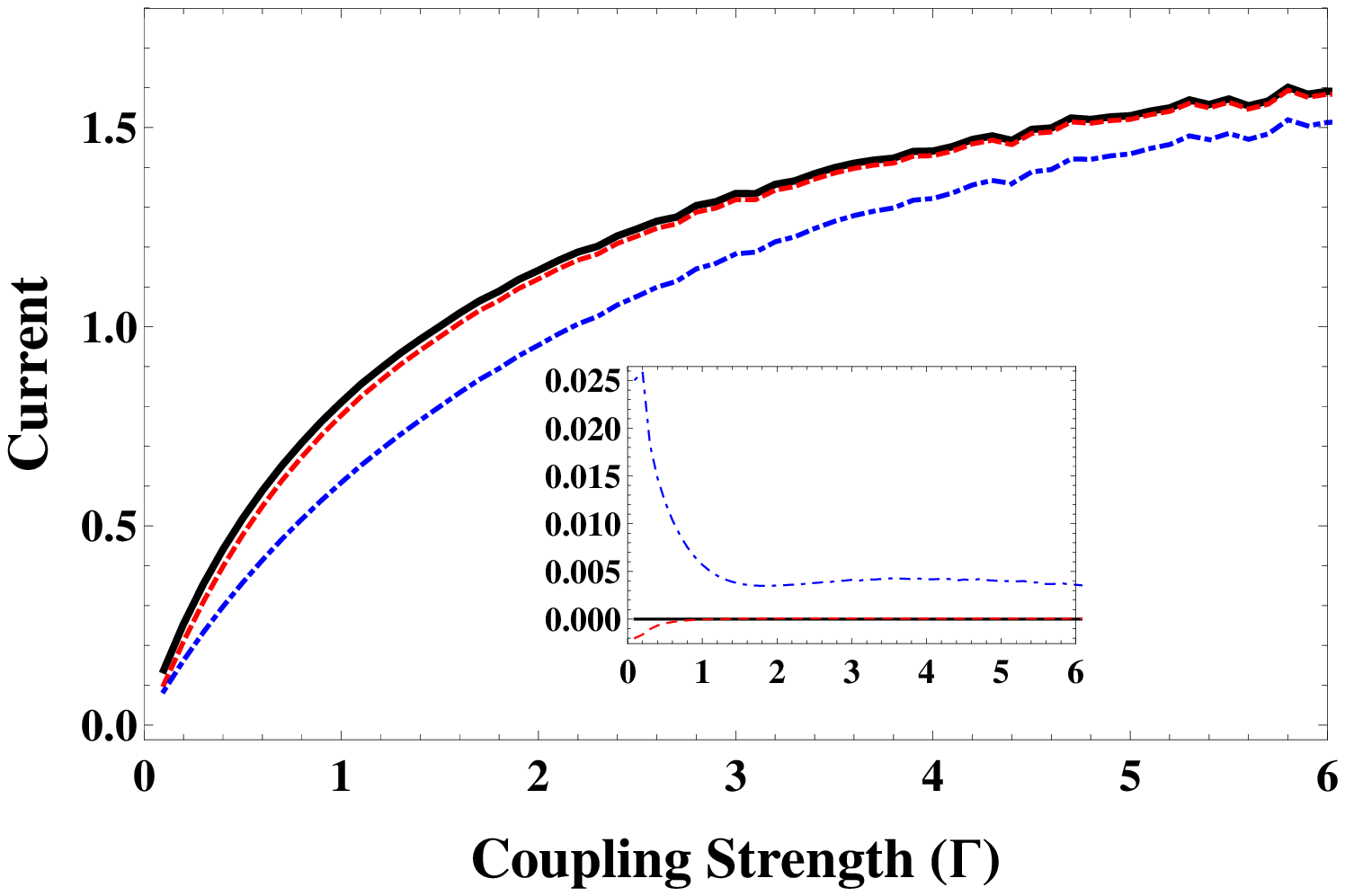} 
\end{tabular}
\caption{(Color online) Net current as a function of $\Gamma$ for serially 
coupled (left) and side coupled (right) double quantum dot system with coulomb interaction treated within HF approximation 
for U=0.0 (continuous), U=0.1 (long dashed) and U=1.0 
(short dashed). 
Here $t=1$, $\beta_L=\beta_R=1000.0$, $\mu_L=\frac{V}{2}$ and 
$\mu_R=-\frac{V}{2}$ with $V=2$. All energy scales are in units of $t$ and 
current is in units of $\frac{e}{h}t$ with $t=1$.
 Difference between current calculated within 
HF and $GW$ approximations for same set of parameters is shown in the 
inset. The wiggles for larger couplings in the right panel are due to 
the numerical errors caused  by a larger grid spacing chosen 
for want of computational time.
}
\label{fig-4}
\end{figure*}

\begin{figure*}[tbh]
\centering
\begin{tabular}{ccc}
\includegraphics[width=5.2cm,height=4cm]{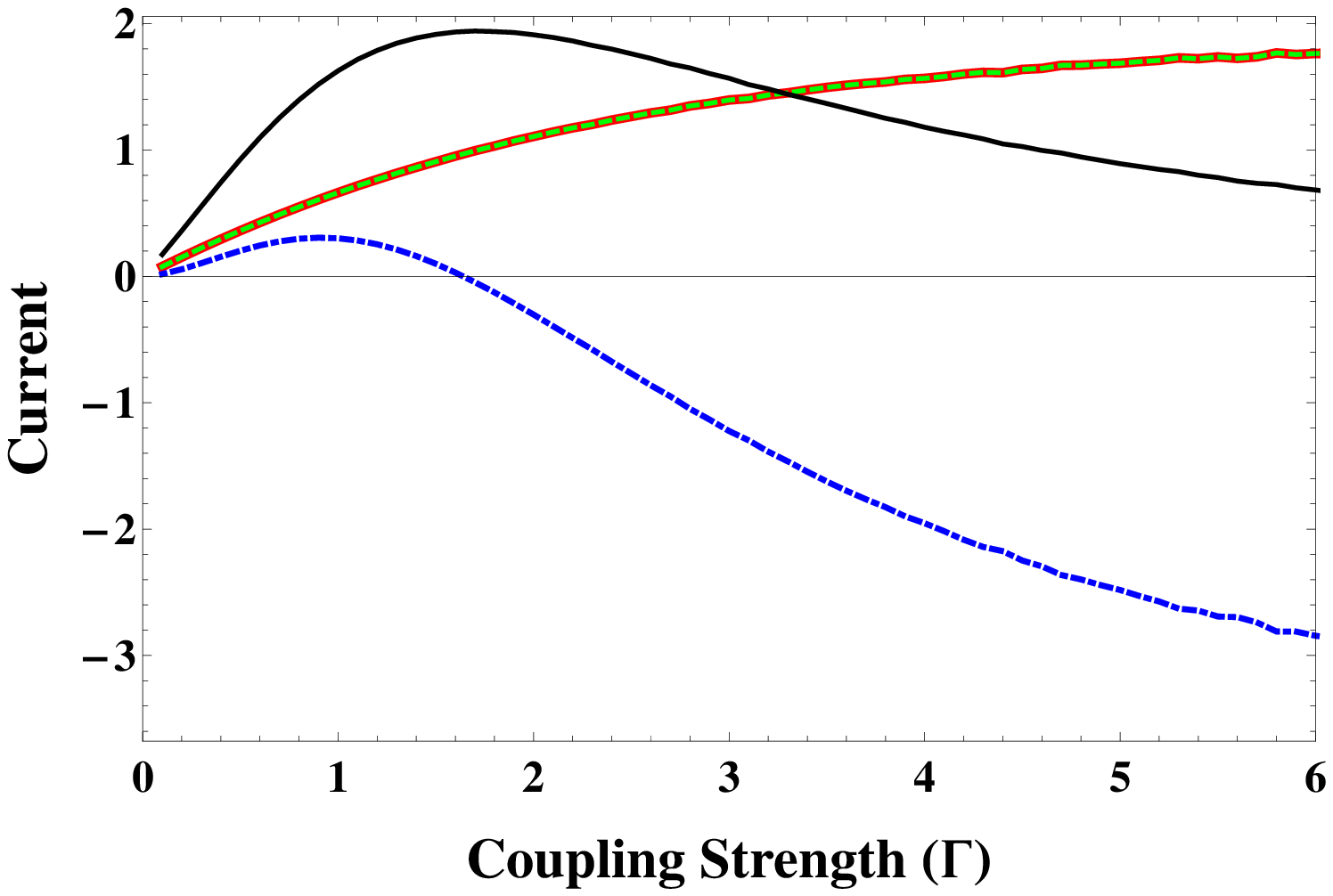} &
\includegraphics[width=5.2cm,height=4cm]{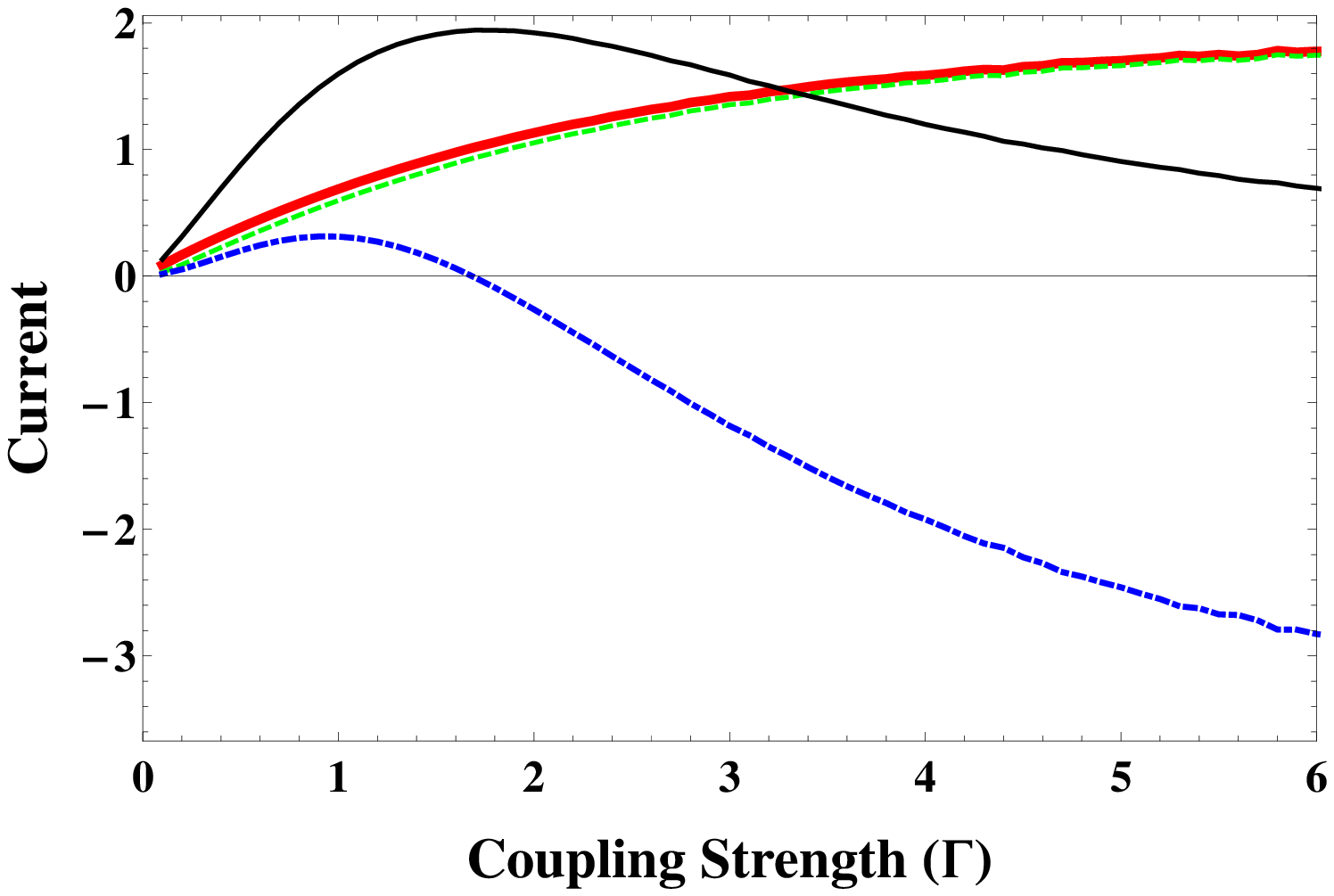} &
\includegraphics[width=5.2cm,height=4cm]{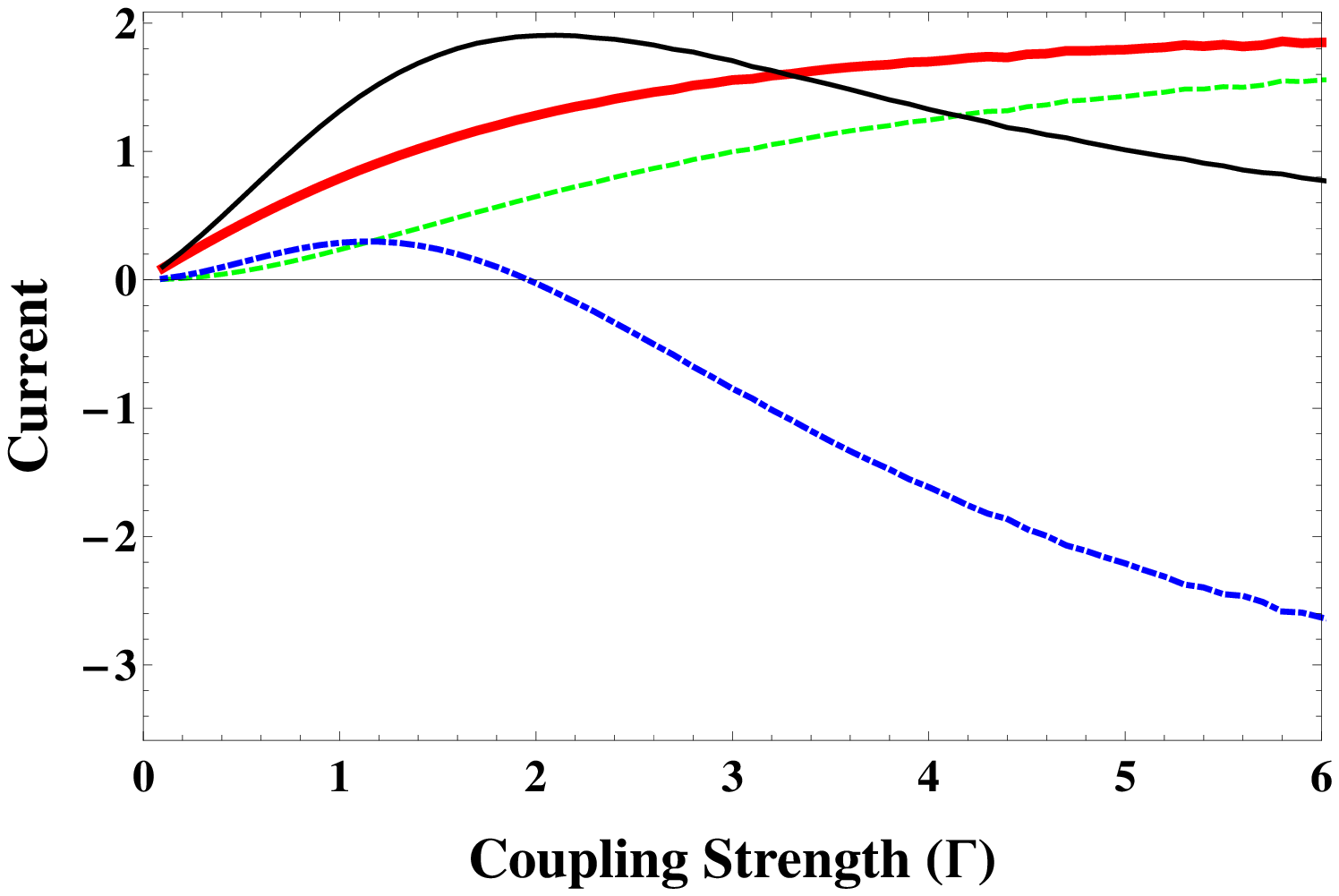} \\ \\
\includegraphics[width=5.2cm,height=4cm]{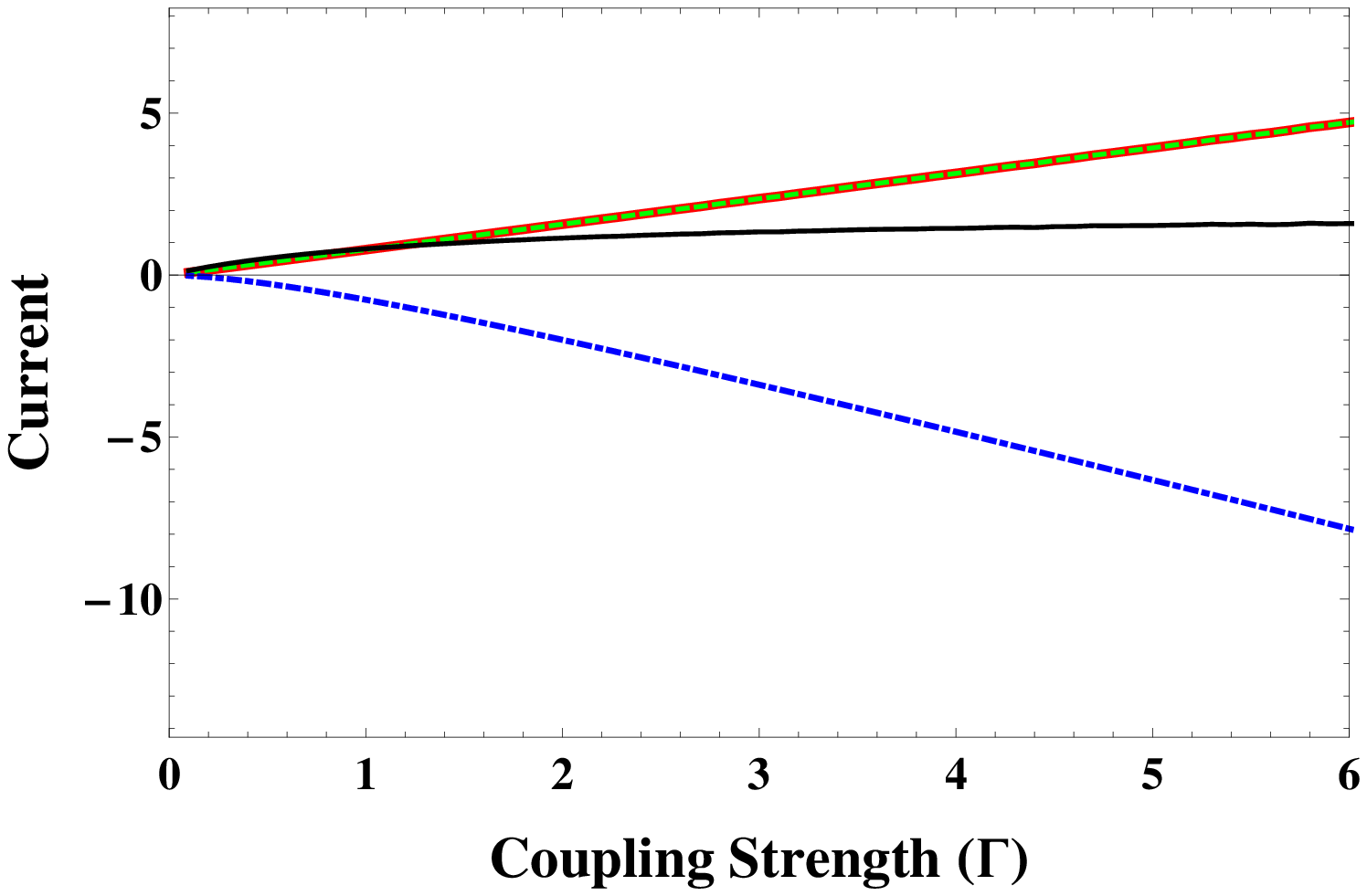} &
\includegraphics[width=5.2cm,height=4cm]{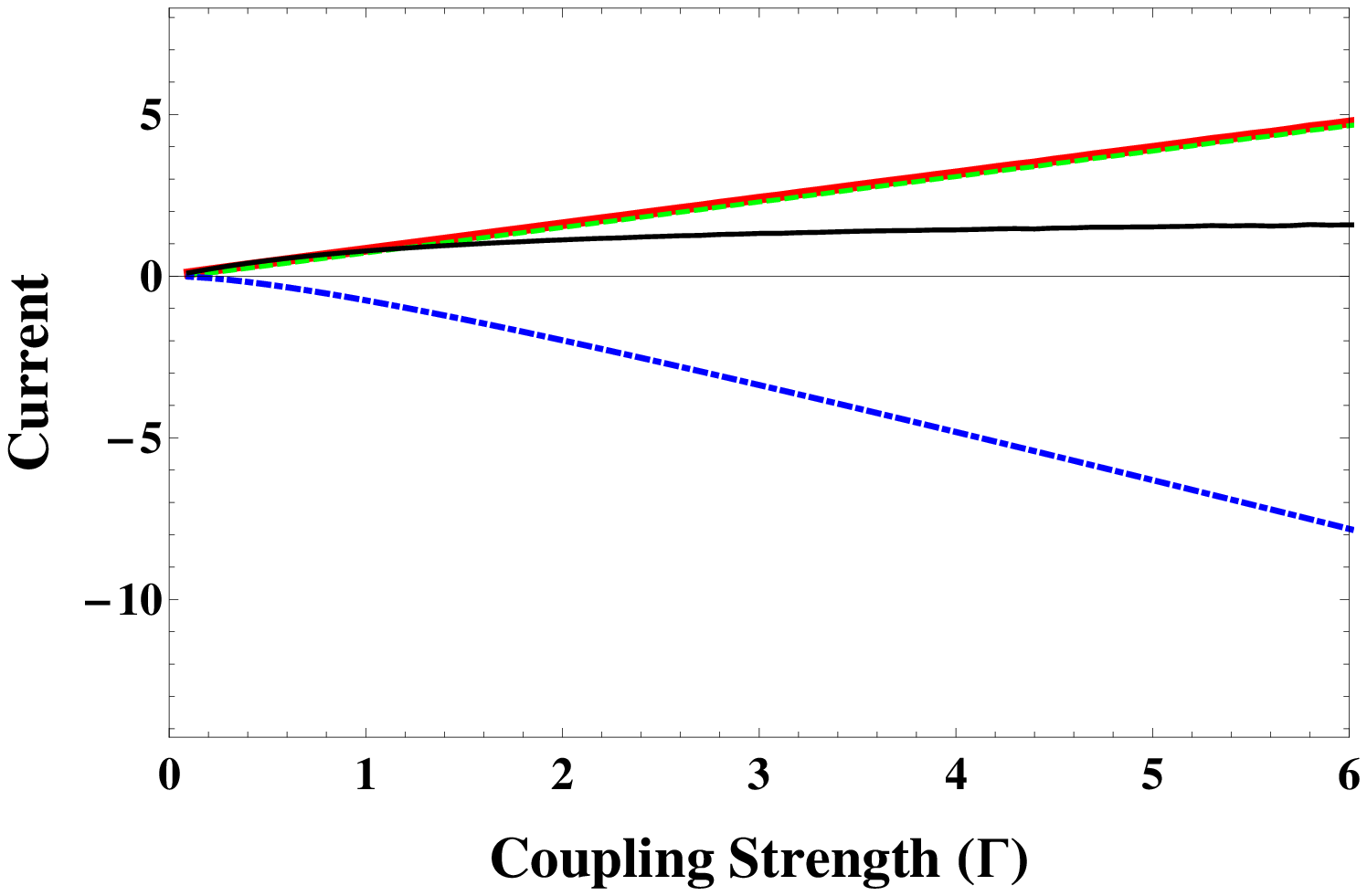} &
\includegraphics[width=5.2cm,height=4cm]{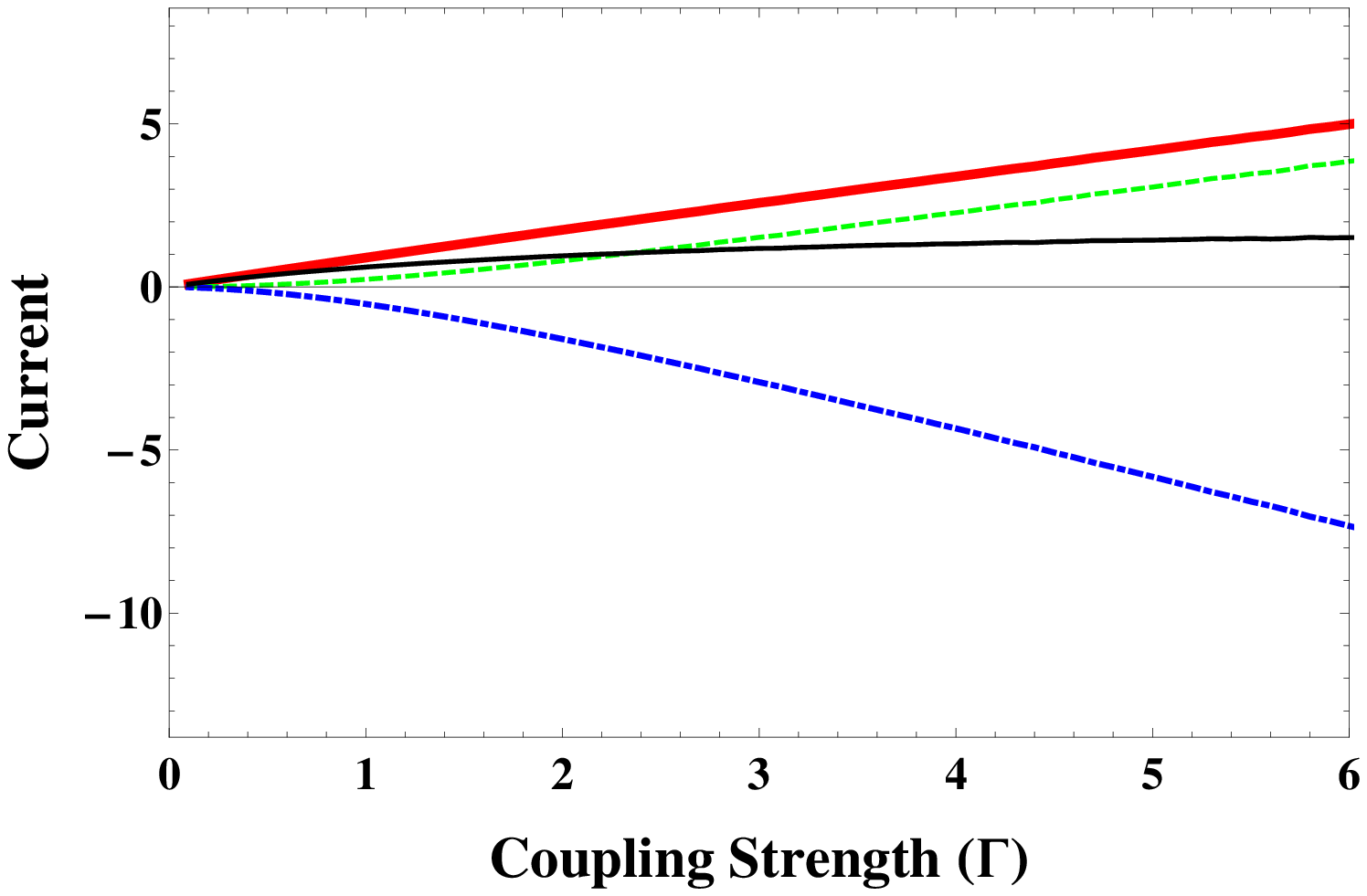} \\
\end{tabular}
\caption{(Color online) Currents carried by bonding (red-thick) state,  
anti-bonding (green-dashed) state, and coherences between them  (blue-dot-dashed) 
along with the net current (black-thin) for serial coupled case (upper-row) and  side 
coupled case (lower-row) with coulomb interaction between electrons on quantum dots treated within HF approximation, 
for U=0.0 (leftmost column), U=0.1 (middle column) and U=1.0 
(rightmost column). Rest of the parameters are same as in Fig. (\ref{fig-4}).
 Note that due to Coulomb interaction, the currents 
carried by the eigenstates are no longer equal.
}
\label{fig-5}
\end{figure*}

The equation of motion for the interacting Green's function together  with self energies $\Sigma_{ij}^{MF}$, 
$\Sigma_{ij}^{GW}$\cite{app} give a set of coupled equations which need to be solved 
self-consistently. This is done most efficiently in the energy domain. The converged
solution of the self-consistent solution for the Greens functions is then used to compute the
net current using Eq. (\ref{eq-8}).
We display the net current as a function of coupling 
strength for various 
Coulomb interaction strengths calculated within HF approximation in Fig. 
(\ref{fig-4}). Inset shows the deviations in the net current calculated  
within the HF approximation from that calculated within $GW$ approximation. 
As can be seen, the previously observed non-monotonic behavior of the 
net current for both the serially-coupled as well as the side-coupled cases are 
robust to weak Coulomb correlations.  However, 
partitioning of the current in terms of the population and coherences,
as introduced previously, is more subtle in the presence 
of interactions. 

At the Hartree-Fock level, if the partitioning is 
done in the single-particle basis renormalized by the mean-field
potential, the individual current components satisfy the left-right
symmetry as discussed previously. However, if the partitioning is done 
in the bare single particle basis (rotation by $\cal U$), the individual 
current components do not satisfy the left-right symmetry in case of 
serially coupled dots. In the side-coupled system, it turns out, that the
left-right symmetry is always satisfied, irrespective of the basis used
(rotation by any arbitrary ${\cal U}$ matrix). Purely on the physical grounds,
we choose to define such a partitioning in the eigenbasis of the renormalized
dots where all current components, both in the serially- and the side-coupled
cases, satisfy the left-right symmetry. In Fig. (\ref{fig-5}), we plot these 
current components as function of the coupling strength. These show qualitatively
similar behavior as obtained in the previous section for non-interacting
system.

It should be emphasized that within the mean-field approximation, 
the single-particle picture is still valid and one can write down 
an effective single-particle Hamiltonian by renormalizing the 
bare dot energies and couplings. This allows to identify a transformation matrix 
${\cal U}$. Such a single-particle
description breaks down within the $GW$ approximation. Thus identifying
the individual current components in the eigenbasis is not possible.
In this case, therefore, we analyze the total current which, irrespective
of the basis used, always satisfies the left-right symmetry.  
We find that even within the $GW$ approximation, the net current 
shows similar qualitative behavior with increasing coupling strength
as discussed previously. This result is presented in Fig. (\ref{fig-4}) with a comparison between the HF and the 
$GW$ results.

\section{Conclusion}
\label{sec-4}

In this work we have explored the effects of system-reservoir coupling on 
currents through molecular (quantum dot)  junctions.  It is shown that the  net 
current in a  molecular junction is not always a monotonically increasing 
function of the coupling strength. We  have demonstrated this by considering two 
simple model junctions which are easily realizable in experiments.
For a serially arranged double quantum dot system, the net current behaves 
non-monotonically and goes 
to zero for large $\Gamma$, while for a side-coupled quantum dot system, the 
current increases monotonically and saturates to a finite non-zero value. These 
two different current behaviors originate due to  competition between the 
classical and 
the quantum contributions to the junction conductance. The classical current, 
described in terms of the eigenstate populations, and the quantum contribution, 
that comes from the superposition between the eigenstates, have opposite 
contributions to the net current. The classical part is always positive (flows 
along the applied bias) while the quantum contribution is always negative for 
large couplings. 
For a serially coupled system,  for large couplings ($\Gamma$), the classical 
and the quantum contributions saturate to the same finite value that corresponds 
to the (quantum) conductivity of a perfect channel. The two contributions  
therefore 
tend to cancel each other out completely at large $\Gamma$, leading to the net 
zero current through the junction.
On the other hand, for a side-coupled system, the two contributions  grow 
linearly with $\Gamma$ in opposite directions
with the same rate. This results in the net current saturating to a finite 
value. The coherent contribution in this case is negative for all $\Gamma$ 
values.   

It is to be noted that while for serially coupled system,  the coherent 
contribution can be positive or negative or even vanish depending on the system 
parameters, for a side-coupled system, however, the coherent contribution is 
always negative and is zero only when the net current vanishes. That is, for a 
side-coupled system, the coherent channel always conducts in the direction 
opposite to the applied bias and can not be blocked to maximize the net current, 
which is possible for a serially coupled system. 
We found that the qualitative results remain valid even for more
realistic junctions with Coulomb interactions.  

In order to preserve the left-right symmetry, the partitioning of the current requires a
careful choice of the basis. We found that partitioning in the eigenbasis 
satisfies this criteria, although it does not rule out possibilities of
other basis. Within HF approximation, since the single-particle picture is still valid,
it is easy to identify the eigenbasis and analyze the current components.
Whereas within the $GW$ approximation, the single-particle picture is not valid.

Although not discussed here, we observed that the qualitative
behavior of the net current with reservoir couplings as discussed here 
remains valid even for reservoirs with more general spectral functions (without wide-band 
approximation). For example, for a Lorentzian bath spectral density, the results 
for large couplings $(\Gamma>2t)$ remain valid for both
model systems. Similarly, for a circular molecular junction, for example 
symmetric four site cyclic molecular junction, the population and coherences 
contribute oppositely to the net current for large coupling strengths. This,
therefore, seems to be a general trend for currents in molecular junctions. 
In fact, for a noninteracting electron system, in general, the classical (population) 
contribution is always positive\cite{app}. The quantum 
(coherence) contribution, as discussed above, can be positive or negative for 
small reservoir couplings but become negative for large coupling strengths.

\section*{Acknowledgments}
H. Y. and U. H. acknowledge the financial support from the Indian Institute of 
Science, Bangalore, India.

\section{Appendix}
\label{appendix}
\subsection{Greens function and its equation of motion}
Green's functions (in matrix form) are defined on Schwinger-Keldysh contour 
\cite{Rammer2007} as,
\begin{eqnarray}
\label{eq-9}
 &&G^{c}(\tau,\tau')=\nonumber\\
 &&-\frac{i}{\hbar}\langle\big[\Theta(\tau,
\tau')\Psi(\tau)\Psi^\dag(\tau')-\Theta(\tau',
\tau)\Psi^\dag(\tau')^T\Psi(\tau)^T\big]\rangle
\end{eqnarray}
where $\tau$ and $\tau'$ are contour times with,
 $\Psi(\tau)= (c_1(\tau), c_2(\tau))^{T}$
and $\Theta(\tau,\tau')$ is the Heaviside step function defined on the
Schwinger-Keldysh contour.
$G^{c}(\tau,\tau')$ satisfies the following equation of motion 
\begin{eqnarray}
\label{eq-11}
 \int_c d\tau_1 && 
\left[\left(i\hbar\frac{\partial}{\partial\tau}-H_{S}\right)\delta^{c}(\tau,
\tau_1)\right.\nonumber\\
&-&\left.\Sigma_{}^{c}(\tau,\tau_1)\frac{}{}\right]G^{c}(\tau_1,\tau^\prime)
=\delta^{c}(\tau, \tau^\prime)
\end{eqnarray}
where $\Sigma^{c}$ is self-energy due to interaction with the reservoirs and it 
is given as sum of 
self energies due to left and right reservoirs i.e., 
$\Sigma^{c}_{}(\tau,\tau')=\sum_{\alpha=L,R}\Sigma^{c}_{\alpha}(\tau,\tau')$. 
The self energies due to reservoirs  is  given by a $2\times 2$ matrix with  
element ($i,j$) defined as, 
\begin{eqnarray}
[\Sigma_\alpha^c(\tau,\tau^\prime)]_{ij}= {g_{\alpha}^{(i)}}^{*}g_{\alpha}^{(j)}
\sum_{k,k'}^{}G^0_{\alpha k,\alpha k'}(\tau,\tau').
\end{eqnarray}
Here $G^0_{Lk,Lk'}(\tau,\tau')$ and $G^0_{Rk,Rk'}(\tau,\tau')$ are contour
ordered Green's functions for the isolated reservoirs. Equation (\ref{eq-11}) 
can be projected onto 
the real times using Langreth rules to obtain the real-time Green's
functions \cite{Haug2008,Rammer2007}. 
At steady-state all Green's functions become time 
translation invariant and can be handled easily in the energy domain. 

\subsection{Self energies due to coulomb interaction}
Coulomb interaction adds an extra self-energy to the equation of motion given in \ref{eq-11}. 
Within mean-field approximation, the self-energy due to Coulomb interaction is given by, 
\begin{eqnarray} 
\label{SE-MF}
\Sigma_{ij}^{MF}(\tau,\tau')&=&-i\hbar\sum_{k=1,2}V_{ik}^{}G_{kk}^{c}(\tau,
\tau^+)\delta_{ij}^{}\delta^{c}_{}(\tau,\tau')\nonumber\\
 &&+i\hbar V_{ij}G^{c}_{ij}(\tau,\tau^+)\delta_{}^{c}(\tau,\tau'),
\end{eqnarray}
where $\tau^+$ is infinitesimally greater than $\tau$, 
and within $GW$ approximation, the self-energy is obtained as,
\begin{eqnarray}
\label{SE-GW}
\Sigma_{ij}^{GW}(\tau,\tau')&=&-i\hbar\sum_{k=1,2}V_{ik}^{}G_{kk}^{c}(\tau,
\tau^+)\delta_{ij}^{}\delta^{c}_{}(\tau,\tau')\nonumber\\
&&+i\hbar G^{c}_{ij}(\tau,\tau')W^{c}_{ji}(\tau',\tau).
\end{eqnarray}
Here $W^{c}_{}$ is the nonequilibrium screened coulomb interaction which 
satisfies the following Dyson like equation,
\begin{eqnarray}
\label{W}
W^{c}_{ij}(\tau,\tau')&=&V_{ij}^{}(\tau,\tau')+\sum_{k_{1}=1,2}\sum_{k_{2}=1,2}
\int_{c}d\tau_{1}\int_{c}d\tau_{2}\nonumber\\
 && V_{ik_{1}}^{}(\tau,\tau_{1})P_{k_{1}k_{2}}^{c}(\tau_{1},\tau_{2})W^{c}_{k_{2}j}(\tau_{2},\tau')
\end{eqnarray}
where $V_{ij}^{}(\tau,\tau')=V_{ij}^{}\delta_{}^{c}(\tau,\tau')$ and 
$P_{ij}^{c}(\tau,\tau')=-i\hbar G_{ij}^{c}(\tau,\tau')G_{ji}^{c}(\tau',\tau)$ is 
the nonequilibrium polarization function within $GW$ approximation.

\subsection{General result for non-interacting system case}
Current carried from right reservoir to left reservoir by the population in the $n^{th}$ system eigenstate coupled to 
two fermionic reservoirs is given by 
\begin{eqnarray} 
I_{L_{nn}}&=&\int_{-\infty}^{+\infty}d\omega\left[\Sigma_{L_{nn}}^{<}(\omega)G_{
nn}^{>}(\omega)-G_{nn}^{<}(\omega)\Sigma_{L_{nn}}^{>}(\omega)\right].\nonumber\\
\end{eqnarray}
For a noninteracting electron system, this can be expressed in the form
\begin{eqnarray}
 I_{L_{nn}}&=&\int_{-\infty}^{+\infty}d\omega 
T_{nn}(\omega)\left[f_L(\omega)-f_R(\omega)\right]
\end{eqnarray}
where the transmission function of the $n^{th}$ state is 
\begin{eqnarray}
T_{L_{nn}}(\omega)&=&(2\pi)^2\sum_{k_1,k_2}\delta(\omega-\epsilon_{Lk_1}
)\delta(\omega-\epsilon_{Rk_2})
\nonumber\\
&\times&|\sum_{m}g_{Lk_1n}G^{r}_{nm}(\omega)g_{Rk_2m}|^2, 
\end{eqnarray}
which is non-negative for any $\omega$. Hence the population channels (for 
$\beta_L=\beta_R=\beta$ case) always conduct current in the direction of applied 
bias. Similarly, the current carried by the coherences between states $m$ and 
$n$ is given by 
\begin{eqnarray} 
I_{L_{mn}}&=&\int_{-\infty}^{+\infty}d\omega\left[\Sigma_{L_{mn}}^{<}(\omega)G_{
nm}^{>}(\omega)-G_{nm}^{<}(\omega)\Sigma_{L_{mn}}^{>}(\omega)\right.\nonumber\\ 
&+&\left.\Sigma_{L_{nm}}^{<}(\omega)G_{mn}^{>}(\omega)-G_{mn}^{<}(\omega)\Sigma_
{L_{nm}}^{>}(\omega)\right].
\end{eqnarray}
This can be simplified to 
\begin{eqnarray}
 I_{L_{mn}}&=&\int_{-\infty}^{+\infty}d\omega 
T_{mn}(\omega)\left[f_L(\omega)-f_R(\omega)\right]
\end{eqnarray}
where the transmission function,
\begin{eqnarray}
T_{L_{mn}}(\omega)&=&(2\pi)^2\sum_{k_1,k_2}\delta(\omega-\epsilon_{Lk_1}
)\delta(\omega-\epsilon_{Rk_2})\nonumber\\
&&\sum_{p,q}\big[\Gamma_{L_{nm}}(\omega)G^{r}_{mp}(\omega)\Gamma_{R_{pq}}
(\omega)G^{a}_{qn}(\omega)\nonumber\\
&+&\Gamma_{L_{mn}}(\omega)G^{r}_{np}(\omega)\Gamma_{R_{pq}}(\omega)G^{a}_{qm}
(\omega)\big].
\nonumber\\ 
\end{eqnarray}
with  $\Gamma_{\alpha_{mn}}(\omega)=2\pi\sum_{k}g_{\alpha k m}g_{\alpha k n}^{*}
\delta(\omega-\epsilon_{\alpha k})$, need not always be positive (nevertheless 
can be shown to be a real quantity).
 
\subsection{Details of numerical solution of equations of motion within HF and $GW$ approximations}
Steady state equations of motion for system Greens function (with the appropriate self energy 
due to coulomb interaction) are Fourier transformed into energy domain, 
the resulting equations are solved self consistently by discretising the 
energy domain with grid spacing $\Gamma/500$ and grid range 
$\approx (-400*\Gamma+10*U,+400*\Gamma+10*U)$.  
All the energy integrals are numerically evaluated using Simpson's-$1/3$ 
rule \cite{Press1994}. Further, convolutions and correlations encountered within $GW$ 
approximation are calculated using 
fast Fourier transform \cite{Press1994} as implemented in FFTW3 \cite{Frigo2005}. 
Convergence of $GW$ calculations is accelerated using Pulay mixing scheme \cite{Pulay1980} as 
implemented in Ref.\cite{Thygesen2008}.

\end{document}